\documentclass[showpacs,preprintnumbers,amsmath,amssymb,nofootinbib]{revtex4}
\usepackage{graphicx}
\usepackage{dcolumn}
\usepackage{bm,color}

\newcommand{\refeq}[1]{~(\ref{#1})}

\begin{document}

\title{ \textbf{Phenomenology from relativistic L\évy--Schr\"odinger equations:}\\
\textbf{Application to neutrinos}}
\author{Nicola \surname{Cufaro Petroni}}
 \email{cufaro@ba.infn.it}
 \affiliation{Department of Mathematics and TIRES, Bari
 University;\\
 INFN Sezione di Bari, \\
via E.\ Orabona 4, 70125 Bari, Italy}
\author{Modesto Pusterla}
 \email{pusterla@pd.infn.it}
 \affiliation{Department of Physics, Padova
 University;\\
 INFN Sezione di Padova, \\
via F.\ Marzolo 8, 35100 Padova, Italy}

\begin{abstract}
\noindent A close connection between Feynman propagators and a
particular L\évy stochastic process is established. The approach can
be easily applied to the Standard Model $SU_C(3)\times SU_L(2)\times
U(1)$ providing interesting, qualitative results. Quantitative
results, compatible with experimental data, are obtained in the case
of neutrinos.
\end{abstract}


\maketitle

\section{Introduction}\label{intro}

In a previous note~\cite{pusterla} we showed that it is possible to
derive and modify the relativistic Feynman propagator of a free
(forceless) particle (fermion spin $\frac{1}{2}$, boson spin $0$ and
$1$) on the basis of L\évy stochastic processes~\cite{applebaum}. We
adopt here the space-time relativistic approach of Feynman's
propagators (for bosons and fermions) instead of the canonical
Lagrangian-Hamiltonian quantized field theory. The rationale for
this choice is that for the development of our basic ideas the
former alternative is better suited to exhibit the connection
between the propagator of quantum mechanics and the underlying L\évy
processes. More precisely, the relativistic Feynman propagators are
here linked to a dynamical theory based on a particular L\évy
process: a point, already discussed in a previous
paper~\cite{cufaro09}, which is here analyzed thoroughly with the
purpose of deducing its consequences for the basic interactions
among the fundamental constituents, namely quarks, leptons, gluons,
photons and so on.

A stochastic process $X(t),\;t\geq0$ on a probability space
$(\Omega,\mathcal{F},\mathbb{P})$ is a L\évy process if
$X(0)=0,\quad \mathbb{P}$-qo, if it has independent and stationary
increments, and if it is stochastically continuous. To simplify the
notation, in this introduction we will restrict ourselves only to
one-dimensional processes, but the three-dimensional extension is
straightforward and will be adopted in the subsequent sections. It
is well known~\cite{sato,applebaum,cufaro08} that all its laws are
infinitely divisible, but we will be mainly interested in the non
stable (and in particular non Gaussian) case\footnote{A law
$\varphi$ is said to be \emph{infinitely divisible} if for every $n$
it exists a characteristic function $\varphi_n$ such that
$\varphi=\varphi_n^n$; on the other hand it is said to be
\emph{stable} when for every $c>0$ it is always possible to find
$a>0$ and $b\in\mathbf{R}$ such that
$e^{ibu}\varphi(au)=[\varphi(u)]^c$. Every stable law is also
infinitely divisible.}. In this case the characteristic functions of
the process increments are $[\varphi(u)]^{\Delta t/\tau}$ where
$\varphi$ is infinitely divisible, but not stable, and $\tau$ is a
time scale. The transition probability density of a particle moving
from the space-time point 1 to 2 then is
\begin{equation}\label{transpdf}
    p(2|1)=p(x_2,t_2|x_1,t_1)=\frac{1}{2\pi}\int_{-\infty}^{+\infty}du\,[\varphi(u)]^{(t_2-t_1)/\tau}e^{-iu(x_2-x_1)}
\end{equation}
and, in analogy with the non relativistic Wiener case, the Feynman
propagator of a free particle is
\begin{equation}\label{propagator}
    \mathcal{K}(2|1)=\mathcal{K}(x_2,t_2|x_1,t_1)=\frac{1}{2\pi}\int_{-\infty}^{+\infty}du\,[\varphi(u)]^{i(t_2-t_1)/\tau}e^{-iu(x_2-x_1)}
\end{equation}
so that the wave function evolution is
\begin{equation}\label{evolution}
    \psi(x,t)=\int_{-\infty}^{+\infty}dx'\,\mathcal{K}(x,t|x',t')\psi(x',t').
\end{equation}
From\refeq{propagator} and\refeq{evolution} we easily
obtain~\cite{cufaro09}
\begin{equation*}
    i\partial_t\psi=-\frac{1}{\tau}\eta(\partial_x)\psi
\end{equation*}
where $\eta=\log\varphi$ and $\eta(\partial_x)$ is a
pseudodifferential operator with symbol $\eta(u)$ which plays the
role of the infinitesimal generator of the process semigroup
$T_t=e^{t\eta(\partial_x)/\tau}$~\cite{applebaum,cont,taylor,jacob}.

It is well known~\cite{sato,applebaum} that $\varphi$ is infinitely
divisible if and only if $\eta(u)=\log\varphi(u)$ satisfies the
L\évy--Khintchin formula
\begin{equation}\label{LK}
    \eta(u)=i\gamma
    u-\frac{\beta^2u^2}{2}+\int_{\mathbb{R}}\left[e^{iux}-1-iux\,I_{[-1,1]}(x)\right]\,\nu(dx)
\end{equation}
where $\gamma,\beta\in\mathbb{R}$, $I_{[-1,1]}(x)$ is the indicator
of $[-1,1]$, and $\nu(dx)$ is a L\évy measure. Then, in the most
common cases of centered and symmetric laws the equation\refeq{LK}
simplifies in
\begin{equation}\label{LKsymm}
    \eta(u)=-\frac{\beta^2u^2}{2}+\int_{\mathbb{R}}(\cos ux-1)\,\nu(dx)
\end{equation}
and $\eta(u)$ becomes even and real. As a consequence the
corresponding operator $\eta(\partial_x)$ is self--adjoint and acts
on propagators and wave functions according to the
L\évy--Schr\"odinger integro--differential equation
\begin{equation}\label{lseq}
    i\partial_t\psi(x,t)=-\frac{1}{\tau}\,\eta(\partial_x)\psi(x,t)
                             =-\frac{\beta^2}{2\tau}\,\partial^2_x\psi(x,t)
                        -\frac{1}{\tau}\int_{\mathbb{R}}\left[\psi(x+y,t)-\psi(x,t)\right]\,\nu(dy)
\end{equation}
The free equation\refeq{lseq} admits simple stationary solutions: if
we take $\psi(x,t)=e^{-iE_0t/\alpha}\phi(x)$ and
$\alpha=m\beta^2/\tau$, we get
\begin{equation}\label{statsol}
    E_0\phi(x)=-\frac{\alpha^2}{2m}\,\phi''(x)-\frac{\alpha}{\tau}\int_\mathbb{R}[\phi(x+y)-\phi(x)]\,\nu(dy),
\end{equation}
so that for a plane wave $\phi(x)=e^{\pm iux}$, and for  a symmetric
L\évy noise, from\refeq{LKsymm} we have
\begin{equation*}
  E_0\phi(x) =-\frac{\alpha}{\tau}\left[-\frac{\beta^2u^2}{2}+\int_\mathbb{R}\left(e^{\pm iuy}-1\right)\,\nu(dy)\right]e^{\pm iux}
           =-\frac{\alpha}{\tau}\left[-\frac{\beta^2u^2}{2}+\int_\mathbb{R}(\cos uy-1)\,\nu(dy)\right]\phi(x)
           =-\frac{\alpha}{\tau}\eta(u)\phi(x)
\end{equation*}
which entails $E_0=-\alpha\eta(u)/\tau$. Then, switching back to
three dimensions, and introducing the momentum $\bm p=\alpha \bm u$,
we obtain~\cite{cufaro09} the relevant equation
\begin{equation}\label{Ep}
    E_0=-\frac{\alpha}{\tau}\,\eta(\bm u)=-\frac{\alpha}{\tau}\,\eta\left(\frac{\bm p}{\alpha}\right)
\end{equation}
which gives the model for the energy-momentum relations that we will
use in the following.

\section{Extended relativistic quantum equations}\label{rqm}

Starting from\refeq{Ep} the particular non stable law (for details
see for example~\cite{applebaum})
\begin{equation}\label{rqmeta}
    \eta(\bm u)=1-\sqrt{1+a^2\bm u^2}
\end{equation}
(from now on we will write $\bm u^2$ instead of $|\bm u^2|$) with
the following identification of the parameters
\begin{equation*}
    a=\frac{\hbar}{mc}\qquad\quad \alpha=\hbar \qquad\quad\tau=\frac{\hbar}{mc^2}
\end{equation*}
will lead to the formula
\begin{equation}\label{relEp}
    E_0=-mc^2\eta\left(\frac{\bm p}{\hbar}\right)=E-mc^2=\sqrt{m^2c^4+\bm p^2c^2}-mc^2
\end{equation}
which is the well--known relativistic kinetic energy for a particle
of mass $m$. The Schr\"odinger equation of a relativistic
free-particle is then easily obtained from\refeq{relEp} by
reinterpreting as usual $E$ and $\bm p$ respectively as the
operators $i\hbar\partial_t$ and $-i\hbar\bm\nabla$
\begin{equation}\label{releq3}
    i\hbar\partial_t\psi(x,t)=\sqrt{m^2c^4-\hbar^2c^2\bm\nabla^2}\,\psi(x,t)
\end{equation}
It has been shown~\cite{applebaum,ichinose1,ichinose2} that the
L\évy process with logarithmic characteristic\refeq{rqmeta} which is
behind the equation\refeq{releq3} is a pure jump process with an
absolutely continuous L\évy measure $\nu(d^3\bm x)=W(\bm x)\,d^3\bm
x$, where~\cite{applebaum,cufaro09}
\begin{equation}
    W(\bm x)=\frac{1}{2a\pi^2|\bm x|^2}\,K_2\left(\frac{|\bm x|}{a}\right)=\frac{mc}{2\hbar\pi^2|\bm x|^2}\,K_2\left(\frac{mc}{\hbar}\,|\bm x|\right)
\end{equation}
($K_\nu$ are the modified Bessel functions~\cite{abramowitz}), so
that from\refeq{symmlseq} the equation\refeq{releq3} takes the
integro-differential form
\begin{equation}\label{intdiff3}
    i\hbar\partial_t\psi(\bm x,t)
    =-mc^2\int_{\mathbb{R}^3}\frac{\psi(\bm x+\bm y,t)-\psi(\bm x,t)}{2\pi^2\bm y^2}\,\frac{mc}{\hbar}K_2\left(\frac{mc}{\hbar}\,|\bm y|\right)d^3\bm
    y
\end{equation}
Of course from\refeq{releq3} one also derives the free Klein--Gordon
and Dirac equations both for scalar, and for spinor wave
functions~\cite{bjorken}
\begin{eqnarray}
  \left(\square-\frac{m^2c^2}{\hbar^2}\right)\psi &=& 0, \label{KG}\\
  \left(i\gamma_\mu\partial^\mu-\frac{mc}{\hbar}\right)\psi  &=& 0.\label{D}
\end{eqnarray}
while the respective propagators satisfy the inhomogeneous equations
\begin{eqnarray}
  \left(\square_2-m^2\right)\mathcal{K}_{KG}(2|1) &=& \delta^4(2|1)\label{propKG} \\
  \left(i\gamma_\mu\partial_2^\mu-m\right)\mathcal{K}_D(2|1) &=& i\,\delta^4(2|1)\label{propD}
\end{eqnarray}
with  $\hbar=c=1$, and $\delta^4(2|1)=\delta(t_2-t_1)\delta^3(\bm
x_2-\bm x_1)$. Let us finally remark that these relativistic quantum
wave equations have been even recently of particular
interest~\cite{delgado1,delgado2} also in the field of quantum
optical phenomena and of quantum information.


We consider now a class of transformations both of $\eta(\bm u)$
in\refeq{rqmeta}, and of the corresponding relativistic total energy
\begin{equation}\label{etot}
    E(\bm p)=mc^2\sqrt{1+\frac{\bm
    p^2}{m^2c^2}}
\end{equation}
by requiring on the one hand that they preserve the infinite
divisibility (so that our model will always be based on a suitable
L\évy process), and on the other that they modify the free equations
of motion\refeq{KG} and\refeq{D} in such a way that in the
subsequent sections we will be able to eliminate the usual field
theoretical divergencies and obtain a mass spectrum. To this end we
propose to extend the energy-momentum formula\refeq{etot} in the
following way
\begin{equation}\label{etotmod}
    E(\bm p)=mc^2\sqrt{1+\frac{\bm
    p^2}{m^2c^2}+f\left(\frac{p^2}{m^2c^2}\right)}
\end{equation}
where $f$ is a dimensionless, smooth function of the relativistic
scalar $p^2/m^2c^2$ (here $p^2=E^2/c^2-\bm p^2$). Of course this
modification entails that $p^2$ no longer coincides with $m^2c^2$
since the standard energy-momentum relation is now changed into
\begin{equation}\label{pquadromod}
    p^2=\frac{E^2}{c^2}-\bm p^2=m^2c^2+m^2c^2f\left(\frac{p^2}{m^2c^2}\right).
\end{equation}
As we will see in the following, this also implies that the mass no
longer is $m$: it will take instead one or more values depending on
the choice of $f$. As a matter of fact, it could appear preposterous
to introduce a function $f$ of an argument which after all is a
constant (albeit different from 1), but this artifice -- while
keeping a viable connection to a suitable underlying L\évy process
-- will lend us the possibility of having both a mass spectrum, and
a new wave equation when we quantize our classical relations.

To see that, we first remark that\refeq{pquadromod} defines the
total particle energy $E$ only in an implicit form. To find it
explicitly we just rewrite\refeq{pquadromod} as
\begin{equation}\label{pquadromod1}
    g\left(\frac{p^2}{m^2c^2}\right)=\frac{p^2}{m^2c^2}-f\left(\frac{p^2}{m^2c^2}\right)=1.
\end{equation}
with $g(x)=x-f(x)$, so that $x=p^2/m^2c^2$ must be one of the
(possibly many) solutions of $g(x)=x-f(x)=1$. Remark that if we
require that $x=1$ (namely $p^2=m^2c^2$) is a solution, then we must
have $f(1)=0$ and $g(1)=1$. If now $g^{-1}(1)$ represents one of the
said solutions, we will have $p^2/m^2c^2=g^{-1}(1)$ so that
\begin{equation*}
    p^2=\frac{E^2}{c^2}-\bm p^2=m^2c^2g^{-1}(1)
\end{equation*}
which can first of all be interpreted as a simple mass re-scaling,
from $m$ to one of the (possibly many) values $M=m\sqrt{g^{-1}(1)}$.
The new hamiltonian then is
\begin{equation}\label{newhamilt}
    E(\bm p)=\sqrt{m^2c^4g^{-1}(1)+\bm p^2c^2}=Mc^2\sqrt{1+\frac{\bm
    p^2}{M^2c^2}}
\end{equation}
and this mass re-scaling $m\to M$ has as its first straightforward
consequence that the associated logarithmic characteristic $\eta$
underwent little changes, so that it is again trivially infinitely
divisible (albeit with different numerical parameters) and hence
still produces acceptable L\évy processes. But there is more: since
$g^{-1}(1)$ can take several different real and positive values, by
means of our extension\refeq{etotmod} we can introduce a mass
spectrum as we will see in the following.

From the modified energy formula\refeq{etotmod} -- by using the
prescriptions $E\rightarrow i\hbar\partial_t$ and $\bm
p\rightarrow-i\hbar\bm\nabla$ -- one now derives the extended
relativistic Schr\"odinger equation
\begin{equation}\label{releq3mod}
    i\hbar\partial_t\psi(x,t)= mc^2\sqrt{1-\frac{\hbar^2\bm\nabla^2}{m^2c^2}+f\left(\frac{\hbar^2\Box}{m^2c^2}\right)}\,\psi(x,t)
\end{equation}
and then in the usual way one can achieve the modified Klein-Gordon
and Dirac equations ($\hbar=c=1$)
\begin{eqnarray}
   \left[\square-m^2f\left(\frac{\square}{m^2}\right)-m^2\right]\psi &=& 0  \label{modKG}\\
   \left[i\gamma_\mu\partial^\mu-m\sqrt{1+f\left(\frac{\square}{m^2}\right)}\,\right]\psi  &=& 0  \label{modD}
\end{eqnarray}
whereas the corresponding Feynman propagators verify the
inhomogeneous equations
\begin{eqnarray}
   \left[\square_2-m^2f\left(\frac{\square_2}{m^2}\!\right)-m^2\right]\mathcal{K}_{KG}(2|1)&=& \delta^4(2|1)\label{modKGprop} \\
   \left[i\gamma_\mu\partial_2^\mu-m\sqrt{1+f\left(\frac{\square_2}{m^2}\!\right)}\,\right]\mathcal{K}_D(2|1)&=& i\delta^4(2|1)\label{modDprop}
\end{eqnarray}
In momentum space\footnote{The connections between the wave
functions and their propagators of the equations\refeq{modKGprop}
and\refeq{modDprop} are then
\begin{eqnarray*}
  \psi(2) &=& \int[\psi(1)\partial_{1\mu}\mathcal{K}_{KG}(2|1)-\partial_{1\mu}\psi(1)\mathcal{K}_{KG}(2|1)]N^\mu(1)\,d^3V_1 \\
  \psi(2) &=& \int\mathcal{K}_D(2|1) \gamma^\mu N_\mu(1)\psi(1)\,d^3V_1
\end{eqnarray*}
$N_\mu(1)$ being the inward normal to the surface at the point $1$.}
they have the following representation
\begin{eqnarray}
  \mathcal{\mathcal{K}}_{KG}(p^2) &=& \frac{1}{p^2-m^2\left[1+f(p^2/m^2)\right]+i\epsilon}\label{modKGpropmom} \\
  \mathcal{\mathcal{K}}_D(p^2) &=& \frac{1}{\gamma^\mu p_\mu-m\sqrt{1+f(p^2/m^2)}+i\epsilon}\label{modDpropmom}
\end{eqnarray}
We finally remark that, in absence of interaction, the
equations\refeq{modKGprop},\refeq{modDprop},\refeq{modKGpropmom}
and\refeq{modDpropmom} will go back to the well known, usual
formulae when $f(x)\to0$.

\section{Phenomenology}\label{phen}


Let us now consider the Feynman perturbative contributions in
renormalized field theories and focus our attention on the
self-energy terms in QED, QCD and in SM  $SU_C(3)\times
SU_L(2)\times U(1)$. The amplitude for a fermion that propagates
from the vertex $X$ to $Y$, if expanded, looks as follows (the
numerator is not essential for our purposes)
\begin{equation}
    A=A^{(0)}+A^{(1)}+A^{(2)}+\ldots
\end{equation}
with zero order
\begin{equation}
    A^{(0)}=Y\frac{1}{\gamma^\mu p_\mu-m\sqrt{1+f(p^2/m^2)}+i\epsilon}X
\end{equation}
If the fermion emits and reabsorbs a virtual boson with mass $M$
($M=M_Z,M_{W^\pm}$ for weak interactions, $M=0$ for photons or
gluons) we have
\begin{equation}
  A^{(1)} = Y\gamma^\rho\frac{1}{\gamma^\mu p_\mu-m\sqrt{1+f(p^2/m^2)}+i\epsilon} C\frac{1}{\gamma^\nu
   p_\nu-m\sqrt{1+f(p^2/m^2)}+i\epsilon}\gamma_\rho X
\end{equation}
where
\begin{equation}\label{Cfunction}
    C=\int\frac{4\pi g_s^2\,d^{\,4}k}{[(p-k)^2-M^2][\gamma^\mu
    k_\mu-m\sqrt{1+f\left(k^2/m^2\right)}]+i\epsilon}=\tilde{A}(p^2)\,\gamma\cdot p+\tilde{B}(p^2)
\end{equation}
and $g_s$ is the electro-weak renormalized coupling or strong quark
gluon coupling~\cite{infdiv}.

For $M=0$ (gluon renormalized mass, see equation\refeq{Cfunction})
we approximate the exact $A(p^2)$ in a simple way by adding an
infinite number of Feynman graphs that contain only one gluon at any
time:\begin{eqnarray}
  A(p^2) &\simeq& \bar{A}(p^2)=Y\left\{\frac{1}{D}+\frac{1}{D}C\frac{1}{D}+\frac{1}{D}C\frac{1}{D}C\frac{1}{D}\ldots\right\}X=Y\frac{1}{D-C}X\label{approxA}\\
  D &=& \gamma_\mu p^\mu-m\sqrt{1+f(p^2/m^2)}
\end{eqnarray}
All other contributions that we neglect here contain two or more
virtual gluons simultaneously in the intermediate states: they are
supposed to be less relevant starting with terms of order $g^4$.
Hence from equation\refeq{approxA} we may obtain:
\begin{equation}
  \tilde{A}(p^2) = Y\frac{1}{\gamma\cdot p-m\sqrt{1+f}-C}X= Y\frac{1}{\gamma\cdot p-m\sqrt{1+f}-\tilde{A}\,p\,\!\!\!/-\tilde{B}}X
\end{equation}
The search of poles of $\tilde{A}(p^2)\simeq A(p^2)$ leads to the
equation
\begin{equation}
    p^2(1-\tilde{A})^2-m^2\left[\sqrt{1+f(x)}+\frac{\tilde{B}(p^2)}{m}\right]^2=0
\end{equation}

Under the approximation $\tilde{A}(p^2)\simeq \tilde{A}(m^2)$ and
$\tilde{B}(p^2)\simeq \tilde{B}(m^2)$ that follow from the very
reasonable assumption $\tilde{A}\ll1,\;\tilde{B}\ll
m\sqrt{1+f(p^2/m^2)}$ we obtain the equation
\begin{equation}
    x=\frac{p^2}{m^2}=\left(\frac{\sqrt{1+f(x)}+\frac{\tilde{B}(m^2)}{m}}{1-\tilde{A}(m^2)}\right)^2
\end{equation}
which for $g_s\to0,\;\tilde{A},\tilde{B}\to0$ gives back the
classical equation
\begin{equation}\label{classeq}
    \frac{p^2}{m^2}=1+f\left(\frac{p^2}{m^2}\right)\qquad\qquad
    x=1+f(x)\,,\qquad x=\frac{p^2}{m^2}
\end{equation}
Note that equation\refeq{classeq} provides poles of $p^2$ at zero
order (no coupling). More specifically, in order to make the
integral\refeq{Cfunction} finite we can take
\begin{equation}\label{effe}
    f(x)=\lambda_0+\lambda_1x+\lambda_2x^2+\lambda_3x^3
\end{equation}
so that to solve\refeq{classeq} will now mean to find the three
zeros $x_0, x_+$ and $x_-$ of $x-f(x)-1$. We can then write
\begin{equation*}
    x-f(x)-1=-\lambda_3(x-x_0)(x-x_+)(x-x_-)
\end{equation*}
and since we know that $f(1)=0$, on the one hand we have
$\lambda_0=-\lambda_1-\lambda_2-\lambda_3$, and on the other we can
always put $x_0=1$. As a consequence it is easy to see that
\begin{align*}
  &  \frac{\lambda_2}{\lambda_3} = -1-x_+-x_+ \qquad\quad
   \frac{\lambda_1-1}{\lambda_3} = x_++x_-+x_+x_-\qquad\qquad
   x_\pm = \frac{-\lambda_2-\lambda_3\pm\sqrt{\Delta}}{2\lambda_3}\\
  & \qquad\qquad\qquad\qquad\qquad\Delta =
  (\lambda_2-\lambda_3)^2-4\lambda_1\lambda_3-4\lambda_3^2+4\lambda_3
\end{align*}
Then, if following our model the three masses are $m_1=m,\,
m_2=m\sqrt{x_+}$ and $m_3=m\sqrt{x_-}\,$, we finally have
\begin{equation}
    x_0=1\qquad x_+=\frac{m_3^2}{m_1^2}\qquad
    x_-=\frac{m_2^2}{m_1^2}\label{lambda}
\end{equation}

We now remind that the field theories we are considering here (QED,
QCD and $SU_C(3)\times SU_L(2)\times U(1)$) are renormalizable and
consequently the function $f(x)$ can be looked at as a smooth
cut-off that regularizes the perturbative terms. \emph{More
precisely the integral $C$ becomes finite (integrand convergent)
under the assumption that the function $f(x)$ is analytic and
appears either as a series expansion or as a polynomial of third
degree at least}. In the latter case one obtains three poles in the
(zero order) fermionic propagator and obviously three physical
masses for appropriate values of its coefficients. \emph{We may
relate this result with the three families of fundamental particles
(quarks) thus describing the flavour phenomenon}: more explicitly
two propagators (charge $-\frac{1}{3}$ and $+\frac{2}{3}$) for
quarks, one for charged leptons ($e^-$, $\mu^-$, $\tau^-$) and one
for neutrinos ($\nu_e$, $\nu_\mu$, $\nu_\tau$). However the
numerical values of these masses cannot be considered valid if
compared with with experimental data, except for particular cases
where the strong and e.m. interactions are absent (see neutrinos in
the next section).

We now consider the Feymnam propagator for fundamental bosons
(gluons, $W^\pm$, $Z$, Higgs). Again we develop it in perturbation
theory and discover that the most significant contribution that must
be added to the zero order derives from the fermion-antifermion
loops. To be precise one has the zero order propagator that connects
a fermion vertex $\gamma_\mu$ with another $\gamma_\nu$:
$A=A^{(0)}+A^{(1)}+\ldots$
\begin{equation}
    A^{(0)}=\frac{1}{p^2-M^2}\left[g_{\mu\nu}+(\xi^2-1)\frac{p_\mu p_\nu}{p^2-\xi^2M^2}\right]
\end{equation}
where $\xi=1$ in t'Hooft-Feynman gauge, $\xi=\infty$ in unitary
gauge and $\xi=0$ in Landau gauge
\begin{equation}
    A^{(1)}=4\pi
    g^2\int\frac{d^4q}{(2\pi)^4}\mathrm{Tr}\left\{\frac{1}{q-p-m\sqrt{1+f\left(\frac{(q-p)^2}{m^2}\right)}}\gamma_\nu\frac{1}{q-m\sqrt{1+f\left(\frac{q^2}{m^2}\right)}}\gamma_\nu\right\}
\end{equation}

\section{Neutrinos}\label{neutr}

Within the scenario of weak interactions, neutrinos need a special
attention. Indeed they are considered to be massless in the Standard
Model but experiments have shown that they have masses. These masses
are much smaller than those of other fundamental particles.
Neutrinos are produced by weak interactions in definite flavor
states ($\nu_e$, $\nu_\mu$, $\nu_\tau$) which are not stationary
mass (energy) states and one expects mixing and oscillations between
the various flavor states. Both mixing and oscillations are
confirmed by experiments. The oscillations are permanent because
neutrinos do not decay~\cite{modification,ref}.

According to our approach the propagator of a  particle of spin
$\frac{1}{2}$ with mass is given by formula\refeq{modDpropmom} and
in the neutrino case we may deal with an $f(x)$ as a third order
polynomial thus obtaining three physical masses $m_1<m_2<m_3$, in
both cases of Dirac or Majorana neutrinos\footnote{This alternative
has not been established yet on experimental grounds.}. The
Lagrangian of interaction neutrinos/other particles can be divided
into two parts $\mathcal{L}_I^{CC}$ (charged current)
$\mathcal{L}_I^{NC}$ (neutral current):
\begin{equation}
  \mathcal{L}_I^{CC} = -\frac{g}{2\sqrt{2}}\,J_\alpha^{CC}W^\alpha
  \qquad\qquad
  \mathcal{L}_I^{NC} = -\frac{g}{2\cos\theta_W}\,J_\alpha^{NC}Z^\alpha
\end{equation}
$g$ electroweak coupling, $\theta_W$ weak angle, $W^\alpha,
Z^\alpha$ are $W^\pm, Z^0$, and $J_\alpha^{CC},J_\alpha^{NC}$
charged and neutral currents respectively. With the neutrino masses
tending to zero the interactions conserve $L_e, L_\mu, L_\tau$
lepton numbers:
\begin{equation}
    \sum L_e=\hbox{\emph{const}}\qquad\quad\sum L_\mu=\hbox{\emph{const}}\qquad\quad\sum L_\tau=\hbox{\emph{const}}
\end{equation}
with non vanishing masses the neutrino mass term in field theory
does not conserve lepton numbers. We have
\begin{equation}
    \nu_{\ell L}=\sum_iU_{\ell i}\nu_i
\end{equation}
where $\nu_i$ is the field of mass $m_i$ and $U$ is the unitary
mixing matrix.

For neutrinos with definite masses there are two possibilities: if
the total $L=L_e+L_\mu+L_\tau$ is conserved they are Dirac particles
(four component spinors), if there are not any conserved lepton
numbers they are two component Majorana particles\footnote{Processes
in which the total $L$ is conserved like $\mu\to e+\gamma$ or
similar ones are permitted in case of mixing Dirac massive neutrinos
(probabilities much smaller than the experimental upper bounds).}
(no difference between neutrinos and antineutrinos). Neutrino-less
double $\beta$-decay $(A,Z)\to(A,Z+Z)+e^-+e^-$ is forbidden if
massive neutrinos are Dirac particles.

If neutrinos are Dirac particles the mixing parameters are three
rotation angles $\theta_{12}, \theta_{23}, \theta_{13}$ plus a phase
factor $\delta$; if they are Majorana's there are two more phases
$\alpha$ and $\beta$ that are irrelevant for the oscillations and
play a role only in the neutrino-less, double-$\beta$
decay~\cite{modification}. More specifically we have
\begin{equation}
    U=\left(
        \begin{array}{ccc}
          1 & 0 & 0 \\
          0 & C_{23} & S_{23} \\
          0 & -S_{23} & C_{23} \\
        \end{array}
      \right)\cdot
      \left(
        \begin{array}{ccc}
          C_{13} & 0 & S_{13}e^{-i\delta} \\
          0 & 1 & 0 \\
          -S_{13}e^{i\delta} & 0 & C_{23} \\
        \end{array}
      \right)\cdot
      \left(
        \begin{array}{ccc}
          C_{12} & S_{12} & 0 \\
          -S_{12} & C_{12} & 0 \\
          0 & 0 & 1 \\
        \end{array}
      \right)\cdot
      \left(
        \begin{array}{ccc}
          1 & 0 & 0 \\
          0 & e^{i\alpha} & 0 \\
          0 & 0 & e^{i\beta} \\
        \end{array}
      \right)
\end{equation}
where $S_{ij}=\sin\theta_{ij}$ and $C_{ij}=\cos\theta_{ij}$.
According to the previous arguments one easily obtains the
probabilities, for neutrinos, of changing or not changing flavors in
vacuum (see the case of atmospheric neutrinos).

If we reduce our analysis to the transition
$\nu_\mu\rightleftarrows\nu_\tau$ in vacuum we obtain
\begin{equation}
    \bm P(\nu_\mu\to\nu_\tau,L)=|\langle\nu_\tau,L|\nu_\mu,0\rangle|^2=C^2_{23}S^2_{23}\sin^2\frac{p_3-p_2}{2}L\simeq C^2_{23}S^2_{23}\sin^2\frac{\Delta m_{23}^2}{4E}
\end{equation}
with $\Delta m_{23}^2=m_3^2-m_2^2$, and $\bm
P(\nu_\mu\to\nu_\mu,L)=1-\bm P(\nu_\mu\to\nu_\tau,L)$, $L$ being the
distance at which we detect $\nu$.

While the oscillations of neutrinos in the vacuum are a kinematical
phenomenon, in highly dense media it becomes a dynamical phenomenon
because of the interaction neutrino-matter\footnote{All neutrinos
interact with electrons and quarks by neutral currents, but only
electronic neutrinos interact with electrons and quarks via charged
currents.} (electrons, quarks ...) mainly due to $\nu_e$-$e$
interaction. The nuclear fusion reactions in the core of the Sun
produce electron neutrinos in a high density medium. Afterwards
neutrinos cross a decreasing density medium before reaching the
surface of the Sun~\cite{modification}.

One then can describe this situation with a phenomenological
potential $V(r)=\sqrt{2}G_FN_e(r)$, $G_F$ is the Fermi constant,
$N_e(r)$ is the electron density at a distance $r$ from the centre
of the Sun. Accordingly we must deal with neutrino effective masses
and effective mixing angles; they are different from those in the
vacuum. In particular one obtains for the angle $\theta_{12}^m$ the
following formula\footnote{Notice the resonance condition
($\theta_{12}^m=\pi/4$) at the density
\begin{equation*}
    N_e=\frac{\delta_m^2\cos2\theta_{12}}{E2\sqrt{2}G_F}
\end{equation*}}
\begin{equation}
    \tan2\theta_{12}^m=\frac{\delta_m^2\sin2\theta_{12}}{\delta_m^2\cos2\theta_{12}-A}
\end{equation}
with $\delta_m^2=m_2^2-m_1^2,\;A=2\sqrt{2}G_FN_eE$, and $E$ is the
neutrino energy~\cite{modification,ref}.

\textbf{Numerical values}: The function $f(x)$ introduced in this
and in the previous note, in the approximated form of a third-degree
polynomial, permits to calculate masses and other physical
properties of the fundamental particles, namely quarks, neutrinos,
gluons, charged leptons ... The calculation is particularly simple,
qualitatively and quantitatively significant, for the massive
neutrinos ($m_1,m_2,m_3$) from the data~\cite{ref} (whose values
might change) where one can ignore the electro-weak coupling and
deal with the zero order only:
\begin{eqnarray}
    \delta_m^2&=&m_2^2-m_1^2\simeq76.6\pm3.5\;(meV)^2\\
    |\Delta
    m^2|&=&m_3^2-\frac{1}{2}(m_2^2-m_1^2)\simeq2\,380\pm270\;(meV)^2
\end{eqnarray}
Indeed we must consider the function $f(x)$ defined in\refeq{effe}
as a universal function for all weak interactions. Consequently the
zeros of $x-f(x)-1$ should almost coincide with those of the charged
leptons and for the neutrino masses $m_1,m_2,m_3$. More precisely,
if we assume on the basis of universality that the ratios of the
masses w.r.t.\ the lighter one are identical in both cases, we
obtain:
\begin{align}
  & \qquad\qquad\qquad\qquad x_0=1=\frac{m_1^2}{m_1^2}=\frac{m_e^2}{m_e^2}\qquad\quad
  x_-=\frac{m_\mu^2}{m_e^2}=\frac{m_2^2}{m_1^2}\qquad\quad
  x_+=\frac{m_\tau^2}{m_e^2}=\frac{m_3^2}{m_1^2}\\
  & x_-m_1^2-m_1^2\simeq76.6\;(meV)^2\qquad\quad
  m_1\simeq4.235\times10^{-5}\;eV\qquad\quad m_2\simeq6\times10^{-3}\;eV \qquad\quad
  m_3\simeq1.4\times10^{-1}\;eV
\end{align}
From these numbers we get
\begin{equation}
    |\Delta m^2|\simeq2.15\times10^{-2}\;(eV)^2
\end{equation}
In an analogous way on may start from assuming $\Delta m^2\simeq
2.38\times 10^3\, (meV)^2$ on the basis of the analysis of the
data~\cite{ref}, and obtain the following values
\begin{equation}\label{numbers}
    m_1\simeq1.42\times10^{-5}\;eV\qquad\quad m_2\simeq2.928\times10^{-3}\;eV \qquad\quad
  m_3\simeq4.924\times10^{-1}\;eV
\end{equation}
thus obtaining $\delta_m^2\simeq8.57\times10^{-4}\,(eV)^2$. A
similar calculation becomes unrealistic for quarks, and basic bosons
because of the strong coupling that is present in addition to weak
and electromagnetic forces.

\section{Conclusions}

As we proposed, the insertion of an analytic function $f(x)$ (for
$x=p^2/m^2$) into the relativistic energy-momentum formula of the
forceless particle preserves the relation between a certain L\évy
stochastic process and quantum relativistic mechanics as in the
usual case~\cite{pusterla,applebaum}. Furthermore the
phenomenological reduction of $f(x)$ as a third degree polynomial
makes the Feynman graphs convergent at any order (due to
renormalizable theories) and creates three poles (in the
$x$-variable)that may be related qualitatively with three masses,
and consequently with the flavor phenomenon (see details in the
Section~\ref{phen}). Hence we derive two propagators for quarks
(charge $-\frac{1}{3}$ and $+\frac{2}{3}$), one for charged leptons
and one for neutrinos.

As far as neutrinos are concerned we obtain quantitative results
that are almost compatible with the experimental data~\cite{ref}.
This is possible (in neutrino's case) because they couple among
themselves or with other leptons via weak interactions only and
$f(x)$ can be assumed universal within the weak interaction
scenario. Hence we can calculate the neutrino mass spectrum from the
zero order propagator, and ignore the expansion of the renormalized
weak coupling. Impossible to consider only the zero order propagator
for other fundamental particles (such as quarks etc...) because of
their e.m.\ and strong couplings.






\begin{thebibliography}{99}

\bibitem{pusterla} \emph{N.\ Cufaro Petroni and M.\ Pusterla}: Mod.\
Phys.\ Lett.\ A 27 (2012) 1250034

\bibitem{applebaum} \emph{D.\ Applebaum}: \textsc{L\évy processes and Stochastic Calculus}
(Cambridge U.P., 2009).

\bibitem{cufaro09} \emph{N.\ Cufaro Petroni and M.\ Pusterla}, Physica A 388
(2009) 824.

\bibitem{sato} \emph{K.\ Sato}:  \textsc{L\'evy processes and infinitely divisible
distributions} (Cambridge University Press,  1999).

\bibitem{cufaro08} \emph{N.\ Cufaro Petroni}, Physica A 387 (2008).
1875.

\bibitem{cont} \emph{R.\ Cont and P.\ Tankov}: \textsc{Financial Modelling With Jump Processes} (Chapman\&Hall/CRC, Boca Raton, 2004).

\bibitem{taylor} \emph{M.E.\ Taylor}: \textsc{Partial Differential Equations}, Vol I--III (Springer, Berlin, 1996).

\bibitem{jacob} \emph{N.\ Jacob}: \textsc{Pseudo-differential Operators and Markov Processes}, Vol I--III, (Imperial College Press, London, 2001--05).

\bibitem{ichinose1} \emph{T.\ Ichinose and H.\ Tamura}, Comm.\ Math.\ Phys.\ 105 (1986) 239.

\bibitem{ichinose2} \emph{T.\ Ichinose and T.\ Tsuchida}, Forum Math. 5 (1993) 539.

\bibitem{abramowitz} \emph{M.\ Abramowitz and I.\ A.\ Stegun} \textsc{Handbook of Mathematical Functions}
(Dover Publications,  1968).

\bibitem{bjorken} \emph{J.D.\ Bjorken and S.D.\ Drell} \textsc{Relativistic Quantum
Mechanics} (McGraw-Hill, New York, 1964)

\bibitem{delgado1} \emph{A.\ Bermudez, M.A.\ Martin--Delgado and E.\
Solano}, Phys.\ Rev.\ Lett.\ 99 (2007) 123602.

\bibitem{delgado2} \emph{A.\ Bermudez, M.A.\ Martin--Delgado and E.\ Solano}, Phys.\ Rev.\ A 76 (2007) 041801 (R).


\bibitem{infdiv} \emph{R.P.\ Feynman}: \textsc{The Theory of Fundamental
Processes} (Benjamin, New York, 1962) Section 28, p.\ 140-3

\bibitem{modification} \emph{S.M.\ Bilenky}: CERN Proc.\ 1999 \textsc{European
School of High Energy Physics}, pp.\ 187-217

\bibitem{ref} \emph{A.\ Bettini}: Riv.\ N.\ Cim.\ 32 (2009) 311 \\
\emph{G.L.\ Fogli, E.\ Lisi, A.\ Marrone, D.\ Montanino, A.\ Palazzo
and A.M.\ Rotunno}: Phys.\ Rev.\ D 86 (2012) 013012


\end{thebibliography}
\end{document}